\documentclass[3p,times,twocolumn]{elsarticle}
\usepackage{caption}
\usepackage{amsmath,amssymb,amsfonts}
\usepackage[figuresright]{rotating}
\usepackage{algorithmic}
\usepackage{graphicx}
\usepackage{textcomp}
\usepackage{xcolor}
\usepackage[backref]{hyperref}
\usepackage{color}
\usepackage{enumerate}
\usepackage{multirow}
\usepackage{makecell}
\usepackage{amssymb}
\usepackage{float}
 \usepackage[linewidth=1pt]{mdframed}
\usepackage{caption}
\usepackage{subcaption}
\usepackage{booktabs}

\begin{document}
\begin{frontmatter}




\title{SGMFQP: An Ontology-based Swine Gut Microbiota Federated Query Platform}


\author{\small Ying Wang${^{1,2\#}}$, Qin Jiang${^{3\#}}$, Yilin Geng${^2}$, Yuren Hu${^2}$, Yue Tang${^2}$, Jixiang Li${^2}$, Junmei Zhang${^{2}}$, Wolfgang Mayer${^{4}}$, Shanmei Liu${^{2}}$, Hong-Yu Zhang${^{2,5,6}}$, Xianghua Yan${^{1,3}}$*, \small Zaiwen Feng${^{1,2,5,6,7}}$*}

\address{
$^1$State Key Laboratory of Agricultural Microbiology, Huazhong Agricultural University, Wuhan 430070, China\\
$^2$College of Informatics, Huazhong Agricultural University, Wuhan,430070,  China\\
$^3$College of Animal Sciences and Technology, Huazhong Agricultural University, Wuhan 430070, China\\
$^4$Industrial AI Research Centre, University of South Australia, Mawson Lakes, SA 5095, Australia\\
$^5$Hubei Key Laboratory of Agricultural Bioinformatics, Huazhong Agricultural University, Wuhan 430070, China\\
$^6$Key Laboratory of Smart Farming for Agricultural Animals, Ministry of Agriculture and Rural Affairs, Huazhong Agricultural University, Wuhan 430070, China\\
$^7$Macro Agricultural Research Institute, Huazhong Agricultural University, Wuhan 430070, China\\
${^*}$Correspondence: Zaiwen Feng (Zaiwen.Feng@mail.hzau.edu.cn) and Xianghua Yan (xhyan@mail.hzau.edu.cn)\\
${^\#}$These authors contributed to the work equally and should be regarded as co-first authors.
}

\begin{abstract}
Gut microbiota plays a crucial role in modulating pig development and health, and gut microbiota characteristics are associated with differences in feed efficiency.  To answer open questions in feed efficiency analysis, biologists seek to retrieve information across multiple heterogeneous data sources. However, this is error-prone and time-consuming work since the queries can involve a sequence of multiple sub-queries over several databases. 
We present an implementation of an ontology-based Swine Gut Microbiota Federated Query Platform (SGMFQP) that provides a convenient, automated, and efficient query service about swine feeding and gut microbiota. The system is constructed based on a domain-specific Swine Gut Microbiota Ontology (SGMO), which facilitates the construction of queries independent of the actual organization of the data in the individual sources. This process is supported by a template-based query interface. A Datalog$^{+}$-based federated query engine transforms the queries into sub-queries tailored for each individual data source, and an automated workflow orchestration mechanism executes the queries in each source database and consolidates the results. The efficiency of the system is demonstrated on several swine feeding scenarios.
\end{abstract}

\end{frontmatter}



\section{Introduction}\label{IN}
Feed efficiency is one of the most important issues for sustainable pig production. Daily-phase feeding (DPF) is a form of precision feeding that could improve feed efficiency in pigs\cite{jiang1,jiang}. Gut microbiota can regulate host nutrient digestion, absorption, and metabolism\cite{jiang3,jiang4}. However, it remains unknown whether gut microbiota differs between the two feeding strategies and how they influence pig growth and feed efficiency during the feeding phase. This present study \cite{jiang} conducts a biological experiment firstly to get some raw data. In this experiment, 204 Landrace-Yorkshire 
pigs (75 d) were randomly assigned into two treatments with three-phase feeding and daily-phase feeding respectively. Fecal samples were collected and analyzed by 16S rDNA sequencing technology. Raw data about swine gut microbiota of the daily-phase feeding strategy and the three-phase feeding program were obtained at 80, 82, 100, 102, 131, 133, 155, and 180 d of age. A relational database, PGMDB, was built to organize the data in a structured form. To understand the effect of the daily-phase feeding strategy on the
gut microbiota of growing-finishing pigs, biologists need to query multiple databases (i.e., PGMDB, gutMgene\cite{gutmgene}, KEGG\cite{KEGG}) to find the difference in gut microbes and the function of gut microbiota between the two feeding programs. However, it is error-prone and time-consuming work since the queries often involve multiple manual sub-query steps in the right order over
distributed and heterogeneous databases. Therefore, a more automated and efficient approach as well as a platform to  provide convenient and rapid query service are desired to support biologists and avoid manual errors.
Federated databases~\cite{federateddatabase}, a popular data integration method, integrate underlying heterogeneous data scattered in multiple different databases. Federated data access is commonly realized by constructing a virtual global view over real data sources as an intermediary representation for users. A mapping from the intermediary representation to the actual representation in each data source is used to rewrite the user's query formulated on the intermediary representation to the actual query that can be executed over the corresponding databases. This federated data access method enables the users to focus on the domain concepts and questions they are concerned about, rather than paying attention to the complex data access details at each data source.

\begin{figure*}[hbt]
	\flushleft
	\includegraphics[width=16.5cm, height=10cm]{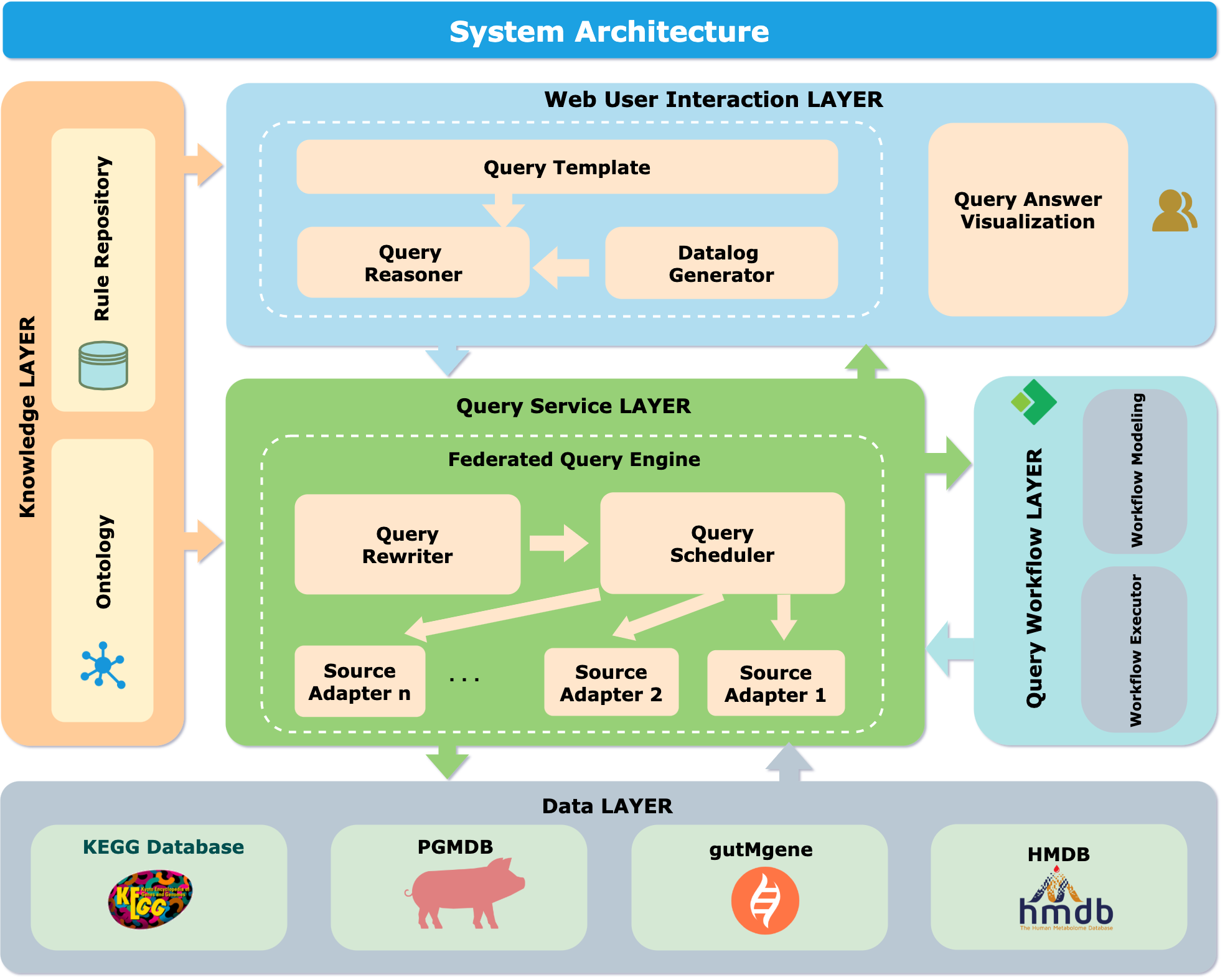}\\
 \caption{Layered architecture of SGMFQP consisting of five layers. The layers include a knowledge layer representing a unified domain model, a data layer involving distributed heterogeneous data sources,  a web user interaction layer providing a template-based user interface, a query service layer implementing the core federated query engine, and a query workflow layer orchestrating the execution of sub-queries.}
\label{system-architecture}
\end{figure*}


\begin{table*}[htb]
 \caption{\newline Queries about swine feeding and gut microbes.}
\centering
\begin{tabular}{|c|c|c|}
\hline

\makebox[0.15\textwidth][c]{\textbf{Query}} & 
\makebox[0.2\textwidth][c]{\textbf{Description}}  \\

\hline 
Q1 & {\makecell[l]{What are the differences in gut microbes and the function of gut microbiota between \\daily-phase and three-phase feeding programs at \textbf{100 d of age} in growing-finishing pigs ?}}\\\hline
Q2 & {\makecell[l]{What are the differences in gut metabolite and the function of gut metabolite between \\daily-phase and three-phase feeding programs at \textbf{155 d of age} ?}}\\\hline
Q3 & {\makecell[l]{What are the differences in the gut microbes and function of the gut microbiota between \\\textbf{180 d and 80 }\textbf{d of age} in growing-finishing pigs?}}\\\hline
Q4 & {\makecell[l]{What are the differences in the gut microbes and function of the gut microbiota between \\\textbf{180 d and 131}\textbf{d of age} in growing-finishing pigs?}}\\\hline
\end{tabular}
\label{queries}
\end{table*}

\begin{table*}[htb]
\caption{\newline OWL 2 constructs used for top-level ontology construction.}
\centering
\scalebox{1} {
\begin{tabular}{|c|c|c|}
\hline
\makebox[0.2\textwidth][c]{\textbf{OWL 2 Feature}} & 
\makebox[0.2\textwidth][c]{\textbf{Meaning}} &
\makebox[0.2\textwidth][c]{\textbf{Example}} \\

\hline
owl:Class & entity type & {\scriptsize \sffamily Swine},{\scriptsize \sffamily Microbiota} \\
\hline
owl:DatatypeProperty & attribute & \textit{gut\_microbiota\_taxonomy\_name}, \textit{microbiota\_id} \\
\hline
owl:ObjectProperty & relationship & \textit{changes\_the\_expression\_by\_microbiota} \\
\hline
rdfs:domain &{\makecell[c]{source class of an\\object property or\\domain of a data \\property}}& {\makecell[c]{The domain of \textit{changes\_the\_expression\_by\_microbiota} \\is {\scriptsize \sffamily Microbiota} or {\scriptsize \sffamily Microbiota} is the domain of \textit{microbiota\_name}}} \\
\hline
rdfs:range & {\makecell[c]{target class of\\an object property}} & {\makecell[c]{The range of \textit{changes\_the\_expression\_by\_microbiota} is {\scriptsize \sffamily Gene}}} \\
\hline
\end{tabular}
}
\label{owl_tags}
\end{table*}

Currently, several important research and contributions have been made in  federated data access. In the field of biology, ontology is used to model and connect different biological resources due to its uniform and unambiguous representation of abstract knowledge concepts
and relationships in a particular domain \cite{domain-ontology}. An ontology-based approach was implemented in \cite{Alejandra2012} to query distributed databases in the cancer domain. They used a federated Local-As-View approach which was not used to provide a unified global view of data sources. BioMart\cite{BioMart}, BioFed\cite{BioFed}, and Bio-SODA\cite{3} use ontology and common nodes or virtual endpoints to connect different data sources. This approach can be difficult to employ if the same data is held using different representations in different databases, which may potentially introduce hard-to-discover errors. KaBOB\cite{KaBOB} and Bio2RDF\cite{Bio2RDF} apply semantic web technology to public databases and create a knowledge repository of linked documents using the Resource Description Framework (RDF) and a common ontology. However, their centralized architecture is difficult to extend and maintain consistency as the underlying sources update.
 
FEDSA\cite{FEDSA} is a data federation platform that proposes a high-level Common Data Model(CDM) and a process-driven data federation method for addressing the challenges in data federation, high-performance processing, etc.  But its customized CDM-based query language is not easy for biologists to use. \cite{Impala} and \cite{Drill} use SQL-like language to express user queries over multiple heterogeneous sources, which can pose difficulty for the biologist who is usually unfamiliar with
the schema of each data source. Therefore, more user-friendly query access is a requirement for domain experts.

OnTop\cite{OnTop} is an open-source system that builds virtual knowledge graphs and implements a transformation from SPARQL queries into SQL queries. However, it is limited to rational databases, failing to federate other heterogeneous data sources. \cite{FAIRVASC} used R2RML\footnote{\href{https://www.w3.org/TR/r2rml/}{https://www.w3.org/TR/r2rml/}} mappings to transform structured data in a relational database to RDF triples. Akin to \cite{Impala}, \cite{OnTop}, and \cite{Drill}, \cite{FAIRVASC} has weak support for heterogeneous data in the storage management mechanism because the non-RDF resources are limited to tabular data. However, it is  necessary to support various data source adaptions since a federated query usually involves multiple data access over the heterogeneous data source.

Query rewriting and reasoning based on ontology and formal rules have been  widely studied to enable the automation and correctness of federated querying. \cite{horn} uses the semantic vocabulary and a context-based unified vocabulary to execute queries across multiple data sources.
\cite{case} discusses data complexity, rewritability, and expressibility problems of Ontology-Mediated Queries (OMQ) formulated in Horn description logic to evaluate the complexity and feasibility of ontology-based data access. In this context, ontology-based federated querying often relies on a query language with formal semantics for reasoning, rewriting, and execution. Datalog\cite{datalog89} as an intermediary query
language has a first-order
logic semantic\cite{first-order-logic} so that it can formally express both queries and
mapping rules precisely to ensure the automation and correctness of the query process. 

In this paper, we present an ontology-based Swine Gut
Microbiota Federated Query Platform (SGMFQP) as a middleware to provide convenient, automated, and efficient query service about swine
feeding and gut microbiota while hiding the underlying heterogeneous data sources. The approach rests on an ontology-mediated federated querying approach that encompasses domain-specific query mechanisms and optimizations.
A Swine Gut Microbiota Ontology (SGMO) is built to provide a unified global query view on top of multiple different data sources. A user-friendly web user interface provides query templates with query reasoning and  visual tabular answers. We use Datalog$^{+}$ to describe formal rules for query reasoning and rewriting. As the core of the SGMFQP, a federated query engine implements query rewriting, query scheduling, and data source adapting and supports sub-query activities in an automatic workflow orchestration\cite{workflow-orchestration}. 
The platform relies on
replication and caching techniques to achieve an efficient query performance. The source code of the system is available at \url{https://github.com/2714222609/fse}.

\section{Materials and Methods}\label{approach}
\subsection{Overview}\label{sec:overview}
In the scenario mentioned in the previous section, the biologist seeks to answer the question \textbf{ `What are the differences in gut microbes and the function of gut microbiota between daily-phase and
three-phase feeding programs at 100 d of age in growing-finishing pigs'}, i.e., \textbf{Q1} in Table \ref{queries}. To answer it, they must query and integrate the information found in multiple databases in a multi-step process: 
\begin{enumerate}[Step 1:] 
	\item Retrieve gut microbes with the difference between daily-phase and three-phase feeding programs at 100 d of age in growing-finishing pigs in the database \verb|PGMDB|. 
	\item Retrieve information on genes affected by these gut microbes in the database \verb|gutMgene|. 
	\item Retrieve pathways in which these genes are involved in the database \verb|KEGG|. 
\end{enumerate}

 The SGMFQP is proposed to provide effective and convenient federated query services to answer the questions about swine gut microbes like \textbf{Q1} involving multiple sub-queries. 
 
 The layered architecture of the SGMFQP is shown in Fig. \ref{system-architecture}. The knowledge layer provides the SGMO ontology and  related rule repository as a medium for ontology-based federated queries across different databases. The data layer involves multiple biological data sources from which the query answers can be obtained. The Web user interaction layer integrates a web-based query template and query generator to express and transform user queries from natural language to formal Datalog$^{+}$ queries. The query reasoner is also integrated with the client to further optimize user queries. The query answers are returned as visual tables with image addresses back to users. The query service layer demonstrates the core of the SGMFQP, a federated query engine, which cooperates with the workflow engine in the query workflow layer, implementing an ontology-based joint query in automatic workflow orchestration. In the federated query engine, the query re-writer implements a mapping from ontology-based user queries to the actual data source. The query scheduler makes a query scheduling plan which decomposes a rewritten Datalog$^{+}$ query into sub-queries. The different source adapters are responsible for translating and executing the respective sub-query in the corresponding data source. The query workflow layer provides a workflow-based model for execution to automate and visualize the query process orchestration.

\begin{figure*}[htb]
\includegraphics[width=1\textwidth,height=0.6\textwidth]{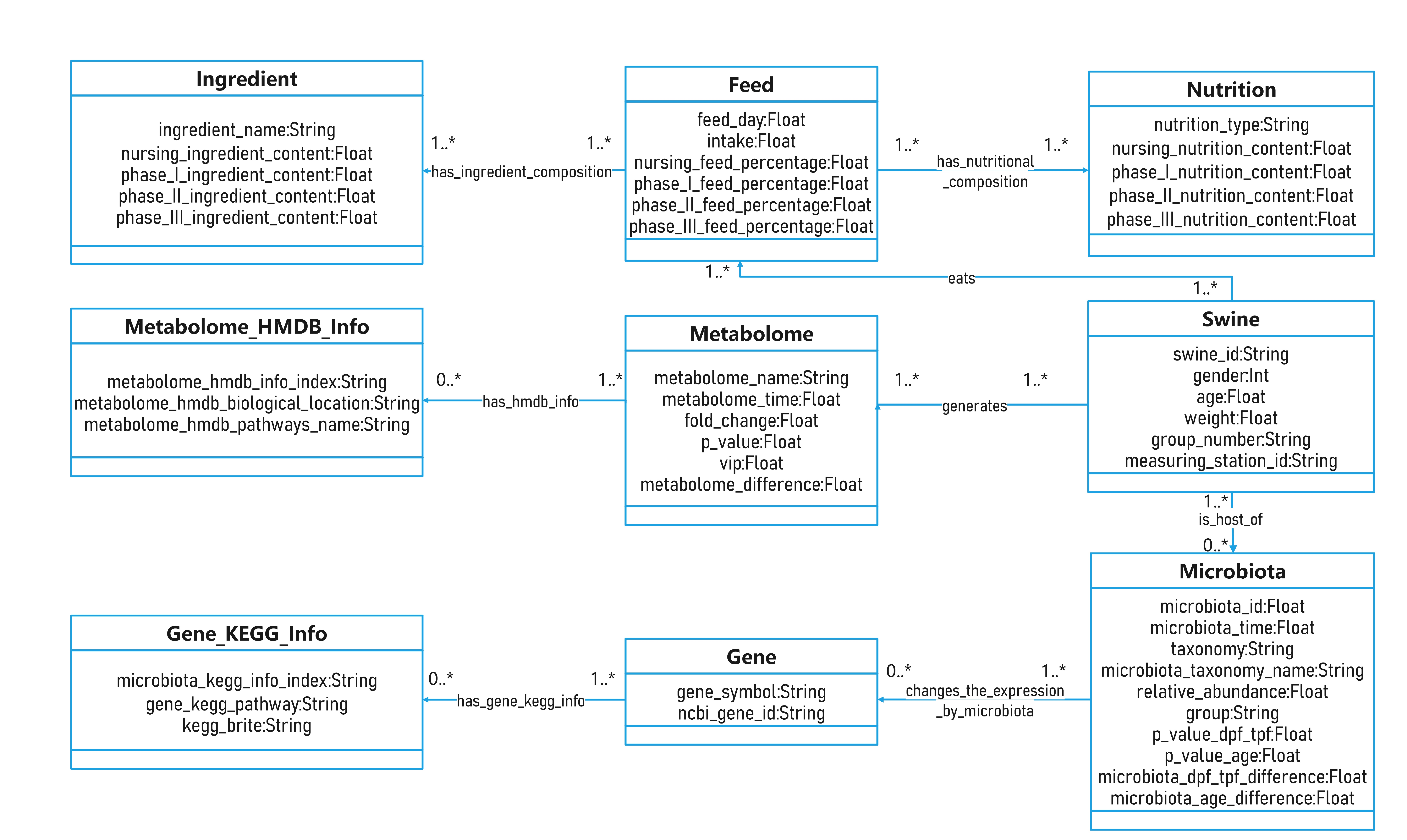}
 \caption{The Swine Gut Microbiota Ontology. Nine classes are constructed to describe feed information, metabolome information, and gut microbiota information of swine. Metabolome information is represented by two classes, Metabolome and Metabolome\_HMDB\_Info. The microbiota information is represented by class Microbiota. Gene information is divided into two parts: Gene and Gene\_KEGG\_Info.}
 \label{ontology}
\end{figure*}

\subsection{Knowledge Layer}\label{sec:knowledge_layer}
Swine Gut Microbiota Ontology (SMGO) is the Ontology Repository constructed by Protégé\cite{protege}.  Table \ref{owl_tags} shows the OWL2 constructs used to define the ontology, including classes, data property, object property, domain, and range constructs. SGMO provides users with a unified domain view that represents the field knowledge of swine gut microbiota and the relationships among them. Users can complete template-based queries without knowing the details of the underlying data source. A graphical representation of SGMO in the Unified Modeling Language (UML\cite{uml}) is shown in Fig. \ref{ontology}, and the Web Ontology Language\footnote{\url{https://www.w3.org/TR/owl2-overview/}} (OWL) description of GMO is freely available at \url{https://github.com/2714222609/fse/blob/master/swine\_gut\_microbiota\_ontology.owl}.


The class of Swine is at the center of the ontology,  which has several attributes including feeding strategy. In this experiment, there are two kinds of feeding strategies, which are the daily-phase feeding strategy and the three-phase feeding strategy. In the data properties of class Metabolome, fold change, P value, and VIP are used to represent the comparison of the metabolome in two strategies. The difference information has been processed in advance and stored in the metabolome\_difference. The data properties of class Microbiota include taxonomy (phylum, family, genus, and species), microbiota\_taxonomy\_name (such as Bacteroidia), microbiota\_dpf\_tpf\_difference (the difference statistics between two strategies at a specific time) and microbiota\_age\_difference (the difference statistics between two different time in a specific strategy). Gut microbiota causes different responses in the body by influencing the expression of host genes. Therefore, the pathways of host genes affected by microbiota are used to represent their functions.

The rule repository consists of Datalog$^{+}$ statements that are used for query reasoning in the web user interaction layer and query rewriting in the query service layer.  Accordingly, the rules are divided into two categories: One expresses the user query constraints at the user level; the other captures the data source mapping rules for mapping the ontology-based user query to the actual data source schemata at the storage level.

\subsection{Data Layer}\label{sec:data_layer}
The data layer involves multiple
heterogeneous data sources distributed in different databases. The desired user query answer is obtained by retrieving information across two or more databases. We considered four main data sources used in our experiment: KEGG (\url{https://www.kegg.jp/}), HMDB (\url{https://hmdb.ca/}), and gutMGene (\url{http://bio-annotation.cn/gutmgene/home.dhtml}), and PGMDB (the raw data were uploaded to the NCBI database \url{https://www.ncbi.nlm.nih.gov/}). KEGG is a non-structured database used to describe the metabolic pathways in which genes are involved. HMDB\cite{hmdb} stores non-structured information about metabolites, and gutMGene shows the effect between the microbes and the expression of those genes. The PGMDB captured our experimental data including the information about pigs, pig feeding, gut microbes, metabolites, difference information between daily-phase and
three-phase feeding programs, etc. The PGMDB was constructed as a relational database using MYSQL (\url{https://www.mysql.com/}). 
The architecture of the SGMFQP can be extended easily to support additional heterogeneous data sources just by providing a corresponding source adapter in the query service layer.

\begin{figure*}[htbp]
  \flushleft
\includegraphics[width=16.5cm, height=10cm]{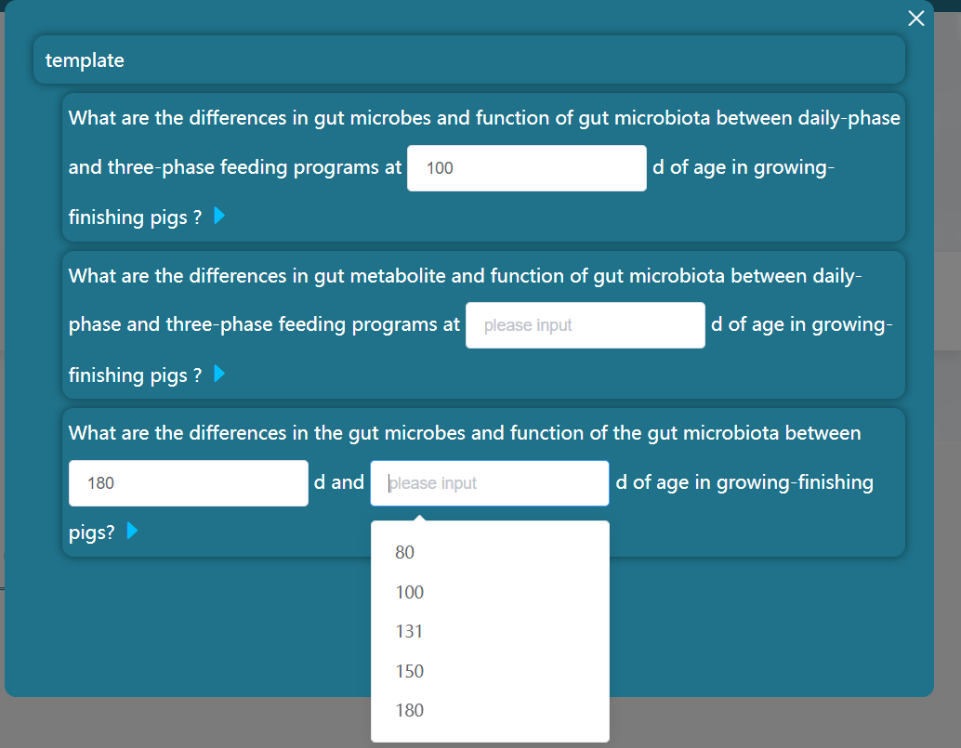}\\
 \caption{Query Template providing natural language-like query expression. This figure shows three types of templates for questions in Table \ref{semantic_rules}. The first template corresponds to \textbf{Q1}. The second template corresponds to \textbf{Q2}. \textbf{Q3} and \textbf{Q4} share the third template.} 
\label{Query Template}
\end{figure*}

\begin{figure*}[htbp]
  \flushleft
\includegraphics[width=16.5cm, height=7cm]{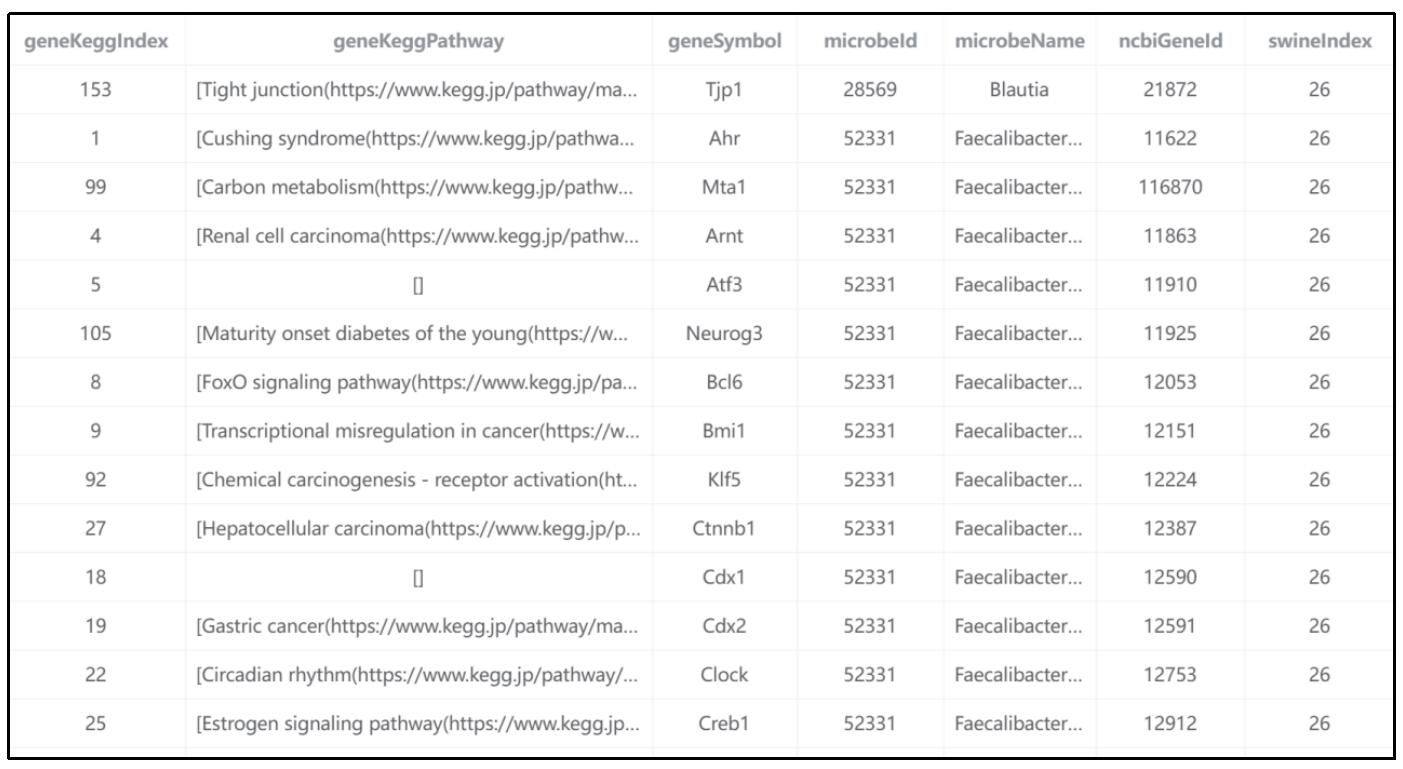}\\
 \caption{Query answer visualization in tabular form. This figure shows the partial result of \textbf{Q1} in {Table \ref{semantic_rules}}.\ \ \ \ \ \ \ \ \ \ \ \ \ \ \ \ \ \ \ \ \ \  \ \ \ \ \ \ \ \ \ \ \ \ \ \ \ \ \ \ \ \ \  \ \ \ \ \ \ \ \ \ \ \ \ \ \ \ \ \ \ \ \ \ \ \ } 
\label{Query Answer visualization}
\end{figure*}
\begin{table*}[htbp]
\caption{\newline FO-based semantic for user requirement rule set divided into four types of rule.}
\label{semantic_rules}
\centering
\begin{tabular}{ccc}
\toprule[1.5pt] 
Rules & Meaning & Formulation \\
\midrule[0.5pt]
 Rule1 & Entity attributes inclusion & \makecell[c]{ $\forall x \exists y (\text { class } ( x ) \rightarrow \text { hasDatatypeProperty } ( x , y ) \wedge$ \\ $\text {DatatypeProperty} (y))$ }\\
 Rule2 & Entity reference of the relationship & $\forall x \forall y(\text { objectProperty } (x,y) \rightarrow \text { Domain} (x) \wedge \text {Range} (y))$\\
 Rule3 & Inverse relationship transformation & $\forall x \forall y (\text { objectProperty } (x,y) \Leftrightarrow \text { InverseObjectProperty} (y,x))$\\
 Rule4 &{Relationship inheritance}&$\forall x \forall y (\text { subObjectProperty } (x,y) \rightarrow \text {objectProperty} (x,y) )$ \\
  \bottomrule[1.5pt]
 \end{tabular}
\end{table*}
\begin{table*}[htbp]
    \caption{\newline The first sub-query Datalog$^{+}$ statements of the \textbf{Q1}  before and after reasoning.}
    \centering
    \begin{tabular}{|c|c|}
    \hline
    \makebox[0.1\textwidth][c]{\textbf{Before Reasoning}} & 
    \makebox[0.1\textwidth][c]{\textbf{After Reasoning}}  \\
    \hline
    {\makecell[l]{    ?(Microbe\_name,Gene\_symbol,Gene\_kegg\_pathway):-\\
    class:Swine(Swine\_index),\\
    class:Microbiota(Microbe\_id),\\
    relationship:is\_host\_of(Swine\_index,Microbe\_id,$<$100$>$),\\
    attribute:p\_value\_dpf\_tpf\_difference(Microbe\_id,$<$1$>$),\\
    attribute:microbe\_name(Microbe\_id,Microbe\_name),\\
    attribute:microbe\_time(Microbe\_id,$<$100$>$).}}
    & {\makecell[l]{    ?(Microbe\_name,Gene\_symbol,Gene\_kegg\_pathway):-\\
    relationship:is\_host\_of(Swine\_index,Microbe\_id,\\$<$100$>$),\\
    attribute:p\_value\_dpf\_tpf\_difference(Microbe\_id,$<$1$>$),\\
    attribute:microbe\_name(Microbe\_id,Microbe\_name),\\
    attribute:microbe\_time(Microbe\_id,$<$100$>$).}}\\
    \hline
    \end{tabular}
    \label{before_after_reasoning}
\end{table*}

\subsection{Web User Interaction Layer}\label{sec:web_layer}
The web user interaction layer provides users with a convenient way to express queries based on predefined templates expressed in natural language text. The queries are transformed automatically to Datalog$^{+}$ queries by the Datalog$^{+}$ query generator, and the queries are further optimized by the query reasoner to reduce query time at the query server layer.

\begin{itemize}
\item \textbf{Query Template}
\end{itemize}
The query template uses the Vue2.0 framework~\cite{vue} to provide a web-based visual query user interface. It pre-defines parameterized templates using the concepts and terms defined in the SGMO. The user query can be expressed as a natural language statement where the user can select values for placeholders as shown in Fig. \ref{Query Template}. The query template mechanism enables the reuse of similar user queries. For example, Q3 and Q4 can share the same template just by instantiating different values. 

 \begin{table*}[htbp]
 \caption{\newline Datalog$^{+}$ query rewriting rules for the first sub-query of the \textbf{Q1}.}
\centering
\scalebox{1} {
\begin{tabular}{|c|c|}
\hline
\makebox[0.2\textwidth][c]{\textbf\ } & 
\makebox[0.1\textwidth][c]{\textbf{Datalog$^{+}$ query rewriting rules}} \\

\hline 
\textbf{relationship} & 
        \makecell[l]{is\_host\_of(X,Y,Z):- :relationship\_entity.is\_host\_of(X,Y,Z).\\} \\
\hline
\textbf{attribute} & 
        \makecell[l]{p\_value\_dpf\_tpf\_difference(X,Y):- :fsmm.microbe(X,X2,X3,X4,X5,X6,X7,X8,Y,X10).\\
        swine\_id(X,Y):- :fsmm.swine(X,Y,X3,X4,X5,X6,X7).\\
        microbe\_name(X,Y):- :fsmm.microbe(X,X2,Y,X4,X5,X6,X7,X8,X9,X10).\\
        microbe\_time(X,Y):- :fsmm.microbe(X,X2,X3,X4,Y,X6,X7,X8,X9,X10).}\\
\hline
\end{tabular}
}
\label{mapping rules1}
\end{table*}

 \begin{table*}[htbp]
 \caption{\newline Rewritten Datalog$^{+}$ statements for the first sub-query of the \textbf{Q1}.}
\centering
\begin{tabular}{|c|c|}
\hline
\makebox[0.11\textwidth][c]{\textbf\ } & 
\makebox[0.1\textwidth][c]{\textbf{Rewritten Datalog$^{+}$ statements}}\\

\hline 
\textbf{relationship} & 
        \makecell[l]{relationship\_entity.is\_host\_of(Swine\_index, Microbe\_id, $<$100$>$).\\} \\
\hline
\textbf{attribute} & 
        \makecell[l]{fsmm.microbe(Microbe\_id,VAR\_1,VAR\_2,VAR\_3,VAR\_4,VAR\_5,VAR\_6,VAR\_7,$<$1$>$,VAR\_9). \\
fsmm.microbe(Microbe\_id,VAR\_1,Microbe\_name,VAR\_3,VAR\_4,VAR\_5,VAR\_6,VAR\_7,VAR\_8,\\VAR\_9).\\
fsmm.microbe(Microbe\_id,VAR\_1,VAR\_2,VAR\_3,$<$100$>$,VAR\_5,VAR\_6,VAR\_7,VAR\_8,VAR\_9).}\\
\hline
\end{tabular}
\label{mapping rules2}
\end{table*}

\begin{itemize}
\item \textbf{Datalog$^{+}$ Query Generator}
\end{itemize}
The Datalog$^{+}$ query generator generates formal Datalog$^{+}$ statements from template-based natural language semi-automatically according to the mapping rules defined in the rule repository at the knowledge layer. For each template schema, the Datalog$^{+}$ query generator searches for the entity, attributes, and relationship between the entities defined in the template, and transforms them to corresponding class, attributes, and relationship skeleton in the Datalog$^{+}$ statement.

\begin{itemize}
\item \textbf{Query Reasoner}
\end{itemize}
The query reasoner is used to optimize user queries for  redundant query removal, inverse query transformation as well as general query refinement based on the four types of rules defined in the rule repository. The query reasoner is integrated into the client rather than the server to further improve the query efficiency of the federated query engine. The reasoning rules are generated automatically from the domain model described by OWL2 according to the rule semantics described by first-order logic as shown in Table \ref{semantic_rules}. 

\textbf{Rule 1} means for any entity who has some data properties, if the entity is a class, then there exists a data property belonging to it. \textbf{Rule 2} indicates an objectProperty relationship between two entities implies that one entity belongs to the domain, and the other entity belongs to the range of the objectProperty. This rule is used to simplify the user query by eliminating redundant conditions in a query. \textbf{Rule 3} shows the equivalence between the objectProperty relationship and its inverse relationship, which can be used to transform the inverse query into an existing forward query. \textbf{Rule 4} describes a relationship inheritance semantic, whereby two entities that satisfy a specific subObjectProperty relationship also satisfy the more general objectProperty relationship. This rule can be used to refine a general query to take into consideration more specific relationships subsumed by the more general relationship.
As shown in Table \ref{before_after_reasoning}, it takes \textbf{Q1} for example, the left part is the first sub-query Datalog$^{+}$ statements of the \textbf{Q1} before reasoning. When applying the \textbf{Rule 2} in Table \ref{semantic_rules}, the Datalog$^{+}$ statement is simplified by removing redundant
Swine and Microbiota classes since they can be deduced
by the relationship is\_host\_of.

\begin{itemize}
\item \textbf{Query Answer Visualization}
\end{itemize}
The user query answers consolidating each sub-query results consists of records in a tabular way, as shown in Fig. \ref{Query Answer visualization}. The column title representing key information is generated according to the concrete user query. Non-tabular information, such as images showing pathways of genes in the KEGG database, are included as links for convenient access as shown in Fig. \ref{Query Answer visualization}.

\subsection{Query Service Layer}\label{sec:query_layer}
The query service is mainly implemented through the federated query engine, including query rewriter, query scheduler, and data source adapter. The query rewriter aims to rewrite the reasoned Datalog$^{+}$ to a form corresponding to the schema of each data source. The query scheduler generates a scheduling plan based on the rewritten Datalog$^{+}$. At the same time, the scheduling plan is handed over to the workflow engine for management. The scheduling plan determines the execution order of each sub-query and the flow direction of the data flow. The data source adapter transforms each sub-query to actual query language executing on the corresponding data source, and consolidate the query results.
\begin{itemize}
\item \textbf{Query Rewriter}
\end{itemize}
The query rewriter uses the rewiring algorithm\cite{re-algorithm} to rewrite the Datalog$^{+}$ queries to a form corresponding to the database schema based on the mapping rules for individual sources, so that the query can be executed on the schema in the source's database. The mapping rules in the query rewriting indicates a corresponding relationship from the domain ontology to actual data source. In Table \ref{mapping rules1}, it defines some rewriting rules of \textbf{Q1}. It can be seen that the \textbf{relationship} in the Datalog$^{+}$ statement is mapped to corresponding relationship table in the database. The \textbf{attributes} are respectively mapped to the corresponding column in the database schema, and other columns are represented 
 by the placeholders with number in sequence (e.g. X3, X4) in the Datalog$^{+}$ statements. After applying the above rules,  the  Datalog$^{+}$ query after reasoning in Table 4 is rewritten in the statements as shown in Table \ref{mapping rules2}. The variables in the rules are all instantiated to concrete column or column value which is aligned with the schema of the data source.




\begin{itemize}
\item \textbf{Query Scheduler}
\end{itemize}
The function of the query scheduler is to identify the rewritten Datalog$^{+}$, and convert the Datalog$^{+}$ into a memory model for storage. According to different questions, the query scheduler can generate different execution plans and workflows to manage the execution process. For example, in \textbf{Q1}, the question involves three distribution sub-queries. First, query the relevant intestinal microorganisms according to the growing and finishing pigs, then query the relevant gene information according to the intestinal microorganisms, and finally query the gene-related metabolic pathways. The query scheduler can parse the Datalog$^{+}$ of the query, generate a scheduling plan according to the three sub-queries involved in the question, correlate the results of each sub-query, and integrate the final query results.

\begin{table}[htbp]
    \caption{\newline SQL statement transformed from the first sub-query of \textbf{Q1}.}
    \centering
   
    \begin{tabular}{|c|}
    \hline
    \textbf{SQL}\\
    \hline
    {\makecell[l]{SELECT swine.swine\_index, \\microbe.microbe\_id,  microbe.microbe\_Name\\
FROM fsmm.swine, fsmm.microbe,\\
     \ \ relationship\_entity.is\_host\_of\\
WHERE is\_host\_of.microbe\_id = \\microbe.microbe\_id\\
  AND is\_host\_of.swine\_index = swine.swine\_index\\
  AND microbe\_dpf\_tpf\_difference = '1'\\
  AND days = '100';}}\\
    \hline
    \end{tabular}
    \label{sql}
\end{table}

\begin{table*}[tbp]
 \caption{\newline Statistical information of the 4 queries.}
\centering
\scalebox{1} {
\begin{tabular}{|c|c|c|c|c|c|}
\hline
\makebox[0.1\textwidth][c]{\textbf{Query}} & 
\makebox[0.2\textwidth][c]{\textbf{Sources}} &
\makebox[0.1\textwidth][c]{\textbf{Results}} &
\makebox[0.2\textwidth][c]{\textbf{System Query(s)}} &
\makebox[0.2\textwidth][c]{\textbf{Direct Query(s)}} \\
\hline
Q1 &KEGG, PGMDB, gutMgene& 128&0.375 & 0.168 \\\hline
Q2 &HMDB, PGMDB, gutMgene& 288&3.632 &  0.453 \\\hline
Q3 &KEGG, PGMDB, gutMgene& 138&0.413 & 0.284 \\\hline
Q4 &KEGG, PGMDB, gutMgene& 119&0.339 & 0.165 \\\hline

\end{tabular}
}
\label{statistic}
\end{table*}

\begin{itemize}
\item \textbf{Source Adaper}
\end{itemize}
The function of the data source adapter is to translate each sub-query into a query statement suitable for each data source. The adapters play a key translation role, as they can parse and execute different sub-queries, and return the query results to the scheduler. In \textbf{Q1}, the PGMDB database and the public KEGG database are involved. When querying the PGMDB database, the Datalog sub-query is translated into an SQL query statement through the MySQL adapter. It shows the SQL statement transformed from Datalog$^{+}$ sub-query of \textbf{Q1} in Table \ref{sql}. The SQL statement is generated automatically by SQL join connection between relationship table and two referenced entity tables. When querying the public KEGG database, the KEGG adapter converts sub-queries into calls to the online RESTful API KEGG provides \cite{restful-api}.

\subsection{Query Workflow Layer}\label{sec:workflow_layer}
\begin{figure}[htbp]
	\includegraphics[width=8cm, height=10cm]{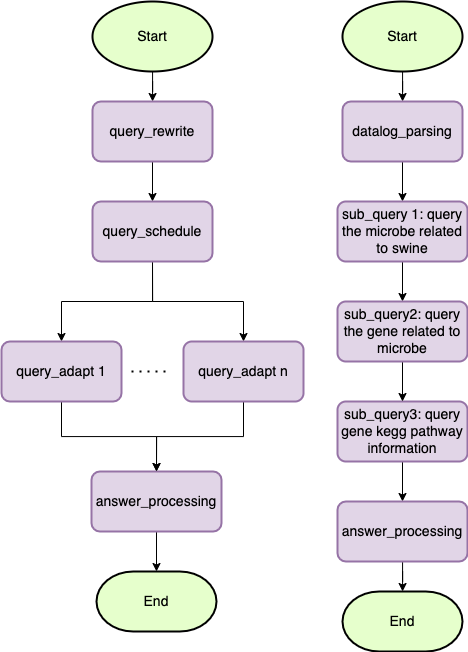}
 \caption{Query workflow model. Left: the core function of the federated service engine, i.e., query rewriting, query scheduling, and query adaptation orchestrated in a workflow. Right: a workflow of Q1 consisting of three sub-queries in sequence.}
\label{workflow}
\end{figure}

The execution process of the scheduling plan is managed by the workflow engine. This system uses the open-source workflow engine Activiti\cite{activiti}, which can generate a set of workflows according to the scheduling plan, and then call different query services according to the execution of the workflow to complete the entire query process. The use process of the workflow includes two steps: workflow modeling and workflow execution.
\begin{itemize}
\item \textbf{Workflow Modeling}
\end{itemize}
After the query scheduler generates the execution plan, the workflow engine will model the execution plan and save the execution process as a BPMN\cite{bpmn} file. BPMN is a general standard language for process modeling, which is used to draw flow charts, save, read and parse them in XML files. It shows the workflow diagram in Fig. \ref{workflow}. On the left is the component at the query service layer orchestrated in a workflow, and on the right is the workflow diagram corresponding to \textbf{Q1} including the process from Datalog$^{+}$ parsing to query answer consolidating.

\begin{itemize}
\item \textbf{Workflow Executor}
\end{itemize}
The workflow executor is responsible for reading BPMN files and executing queries according to the process. During the execution of the workflow, each node of the business process will be read into the database, so that each node (including the start node and end node) is a record in the database. When the query process is executed, the next node will be continuously read from the business process diagram, that is, the automatic management of the process will be realized by operating the database records corresponding to the node.

\section{Results and Discussion}\label{EV}
The performance of the ontology-based query mediation system was tested empirically on several queries in the domain of pig production. This section presents the results of the evaluation of the system. Four specific biological questions in Table \ref{queries} were used as test cases to evaluate the performance of the system.
The experiments were run on an Apple M1 8-core processor computer with 16GB of memory running Mac OS Monterey, and the system was implemented in the Java programming language.

\subsection{Query Execution Performance}
We evaluated the query results and query time of the four questions shown in Table \ref{queries}. Q1, Q3, and Q4 involve the KEGG and PGMDB databases, while Q2 involves the PGMDB and HMDB databases. We also compared the execution times when using the federated query system (SGMFQP) with those obtained when directly querying each source with manually created SQL statements. Table \ref{statistic} shows the number of query results and query time for the four scenarios. The system takes 0.3-0.4 seconds for Q1, Q3, and Q4, and it takes about 3.6 seconds for Q2. In all cases, the query results obtained from the ontology-based mediation approach and that obtained from the direct querying approach were identical, which confirms the accuracy of the system.

\subsection{Comparison Between Local And Online Query}
Different questions include multiple sub-queries, and each sub-query involves different databases. The federated query system provides users with two different query patterns: local query and online query. For local queries, we processed and stored data from different databases (such as KEGG and HMDB) locally based on ontology, built a local database, and the system will directly read data from the local database when querying. The advantages of the local query are short query response time and high query efficiency, while the disadvantages are that the data stored locally is not updated in real-time. The online query method is to use a RESTful API to query the data source. The advantage of this pattern is that the data is obtained online in real time, while the disadvantage is that the query takes longer. The following Table \ref{local and online} shows the comparison of the two query patterns, it can be seen from the table that the online query takes much more time than the local query.
\begin{table}[H]
 \caption{\newline The response time of the local and online query.}
\centering
\scalebox{1} {
\begin{tabular}{|c|c|c|}
\hline
\makebox[0.1\textwidth][c]{\textbf{Query}} & 
\makebox[0.14\textwidth][c]{\textbf{Local Query(s)}} &
\makebox[0.14\textwidth][c]{\textbf{Online Query(s)}} \\
\hline
Q1 & 0.375 & 453.251 \\\hline
Q2 & 3.632 &  388.917 \\\hline
Q3 & 0.413 & 482.325 \\\hline
Q4 & 0.339 & 464.605 \\\hline
\end{tabular}
}
\label{local and online}
\end{table}

\subsection{System Performance With And Without Cache}
To improve the query efficiency, the system introduces Redis as the query cache. Redis is an efficient cache tool, which uses the key-value pattern to cache the query answer. After receiving the Datalog$^{+}$ query, the system calculates its hashcode as the cached key and uses the query answer as the cached value. When the user queries the same problem next time, the system can respond within 0.1 seconds. At the same time, to ensure the timeliness of the cache, we set the expiration time of the Redis cache to 30 seconds. Table \ref{cache} shows the comparison after using the cache. It can be seen that the query time after using the cache is within 0.1 seconds, which we consider to be satisfactory for an interactive querying system.
\begin{table}[htb]
 \caption{\newline The response time of queries without and with caching.}
\centering
\scalebox{1} {
\begin{tabular}{|c|c|c|}
\hline
\makebox[0.1\textwidth][c]{\textbf{Query}} & 
\makebox[0.14\textwidth][c]{\textbf{No Cache (s)}} &
\makebox[0.14\textwidth][c]{\textbf{Cache(s)}} \\
\hline
Q1 & 0.375 & 0.005 \\\hline
Q2 & 3.632 &  0.003 \\\hline
Q3 & 0.413 & 0.004 \\\hline
Q4 & 0.339 & 0.003 \\\hline
\end{tabular}
}
\label{cache}
\end{table}

\section{Conclusion}\label{CL}

To provide a more convenient, automated, and efficient query service about swine feeding and gut microbiota, we presented a Swine Gut Microbiota Federated Query Platform (SGMFQP) based on the SGMO ontology we constructed.
The SGMFQP provides a user-friendly template-based query interface and visual query answers. The core federated query engine implements Datalog$^{+}$ rule-based query rewriting and an orchestration engine executes multiple sub-queries in an automatic workflow. The efficiency of the system is supported by replication and caching techniques and a multi-stage query rewriting approach, as demonstrated on several swine-feeding query scenarios across multiple data sources. 
The layered architecture enables extensibility to additional applications and data sources by adding an application-specific domain ontology and developing additional data source adaptors. The core federated query engine  and workflow-based query process orchestration can be reused in other applications where querying multiple data sources is required.  The SGMFQP will be applied in practical microbiota-based precision feeding of pigs to further optimize feeding through joint analysis of gut microbial composition and function, feed nutrition database, pig dynamic nutrition requirement model across multiple databases on the gut microbiota characteristics of pigs, which will aid in the sustainable development of the pig industry.

In the future, we intend to further improve the characteristics of the system as follows. Currently, the query rewriting rules are constructed manually. An automatic \emph{schema-mapping} method will be proposed to generate these mapping rules. Moreover, maintaining consistent mappings and queries as the underlying data source schemas evolve is another line of future research. Automated generation of a query scheduling plan is an additional concern for future improvement of the system.

\section*{Acknowledgement}
This research project was supported in part by the open funds of the State Key Laboratory of Agricultural Microbiology, Huzhong Agricultural University, and in part by the Fundamental Research Funds for the Chinese Central Universities under Grant 2662020XXQD01, 2662022JC004.

\bibliographystyle{elsarticle-num}
\bibliography{references}

\end{document}